# USING VIRTUAL PARTS TO OPTIMIZE THE METROLOGY PROCESS

WOLFF Valery [1], TRAN Dinh-Tin [2], RAYNAUD Stephane [3]

[1] *IUT Lyon 1, University of Lyon, France, e-mail: valery.wolff@univ-lyon1.fr*
[2] *LamCos INSA, University of Lyon, France, e-mail: dinh-tin.tran@insa-lyon.fr*
[3] *MIP2 INSA, University of Lyon, France, e-mail: stephane.raynaud@insa-lyon.fr*

**Abstract:** *In the measurement process, there are many parameters affecting the measurement results: the influence of the probe system [3,5,7,15,10,16,17,19,21], material stiffness of measured workpiece [7], the calibration of the probe with a reference sphere [3], the thermal effects [14]. We want to obtain the limits of a measurement methodology to be able to validate a result. The study is applied to a simple part. We observe the dispersion of the position of different drilled holes (XYZ values in a coordinate system) when we change the quality of the part and the method of calculation. We use the Design of Experiment [12] (Taguchi method) to realize our study. We study the influence of the part quality on a measurement results. We consider two parameters to define the part quality (flatness and perpendicularity). We will also study the influence of different methods of calculation to determine the coordinate system. We can use two options in Metrolog XG software (tangent plane with or without orientation constraint). The originality of this paper is that we present a method for the design of experiment that uses CATIA (CAD system) to generate the measured parts. In this way we can realize a design of experiment with a largest number of experimental results. This is a positive point for a statistical analysis. We are also free to define the parts we want to study without manufacturing difficulties.*

## 1. Introduction

The control of technical parts is very important in many industries, for example: the plastic industry, aviation, automotive or electronics. Currently, measured parts of mechanical systems are increasingly demanding higher accuracy, which is why the instruments are continuously improved (CMM, optical measuring systems without contact, ...). The measurement methods for a same part to control can be various. You can choose classical (calliper, ...) or high-technology material (CMM, Vision, ...). The best result is not always given by the most expensive and the most accurate machine. It is also linked to the way of calculation proposed by the measurement system software.

In our study, we want to give a help to define the best way to control a part. This work is a first step to reach this goal. We want to apply different methods of calculation to a same measurement. In this way, we hope to find a significant effect of the method of calculation on the result.

## 2. Problematic

In this paper, we are working with a simple part (see fig. 1): a drilled hole on a face, and a geometrical specification of location with a datum coordinate system. In a perfect world, the all faces of the part are perfectly orthogonal, the drilled hole is a perfect perpendicular cylinder and all plane surfaces are without any default of form. In this case, a lot of calculation methods will give the same result: barycentric method, least square optimization, Tchebychev method [13] But a real part as some defaults: flatness, cylindricity, perpendicularity of the faces, and perpendicularity of the axis.

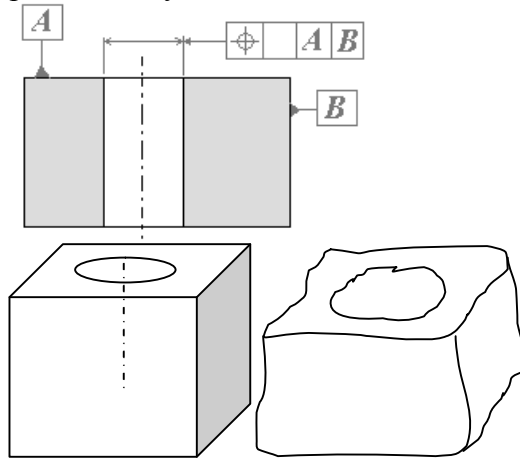

**Figure 1:** *Simple part for experiment: theoretical and skin model*

The software we use with the CMM (coordinate measuring machine) gives different options (see fig. 2 and fig. 3). We can distinguish the nominal element (a), the skin model element (b) and the associated element. The association of a theoretical element to a surface can be made by different ways. The operator can choose to use the option "tangent" or not. It gives different results: theoretical plane associated to a surface by the least square method (c) or the tangent plane associated with the same method (d) [20].

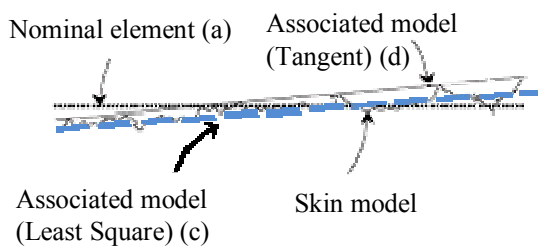

**Figure 2:** *Different models for the association*

The possibilities of the software are also giving different ways to calculate the datum coordinate system. It is possible to use a constraint when we calculate an associated plane. For example, to be perpendicular to a first plane or to pass through a specific point...

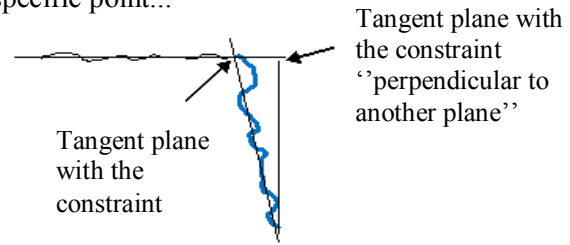

**Figure 3:** *Different constructions of a plane*

The Fig. 3 presents two ways for the construction of a tangent plane in a datum plane system. The difference is by adding a constraint to the plane before to calculate it.

## 3 The study

The study is limited to a simple part. The reason is that we use a method to construct parts. In this way we are able to observe the effects of some parameters on the result of the measurement. In a classical method, we would have machined a lot of different parts. The difficulty to machine an "exact" part with the default we want is erased by our methodology. We construct virtual parts on CAD system (CATIA), and we can have exactly the defaults we want for each part of the study.

### 3.1 The design of experiment (DOE)

The design of experiment is applying the method of Taguchi Design of Experiment [6, 9, 10, 20]. This is a way to reduce the number of experiments to find the effect of a parameter on a result. Documentation [1, 18] gives an array with different tables of Taguchi. If we look in the "array selector" (Table 1), we can obtain different tables able to be used, if you have defined the number of parameters and the different levels that the parameters can take [4,11].

| | | Numbers of parameters | | | | | | | |
|---|---|---|---|---|---|---|---|---|---|
| | | 2 | 3 | 4 | 5 | 6 | 7 | 8 | 9 | 10 |
| Number of levels | 2 | L4 | L4 | L8 | L8 | L8 | L8 | L12 | L12 | L12 |
| | 3 | L9 | L9 | L9 | L9 | L18 | L18 | L18 | L27 | L27 |
| | 4 | L16 | L16 | L16 | L16 | L32 | L32 | L32 | L32 | L32 |
| | 5 | L25 | L25 | L25 | L25 | L25 | L50 | L50 | L50 | L50 |

**Table 1:** *Choosing a Taguchi L(i) table*

*3.1.1 The parameters of the DOE*

We choose to make the experiments with 3 parameters, with each 3 levels. The first parameter is the quality of the top plane. The second is the quality of the side plane. The third parameter is the angle between the two planes. For this first study, we didn't consider the defaults of the measured cylinder (see fig. 4).

The parameter of flatness for the top and the side plane of the part are defined by the machining process used to obtain the surface. The 3 levels we use are: rough milling, finishing milling, and grinding. The values are given in the same order: 0,03 mm, 0,006 mm, and 0,0015 mm flatness default.

According to the document (table 1), we select the table of the Taguchi L9 experiment [12]. That means that we will perform 9 experiments to examine the influence of all parameters to the result of measuring the position of a cylinder.

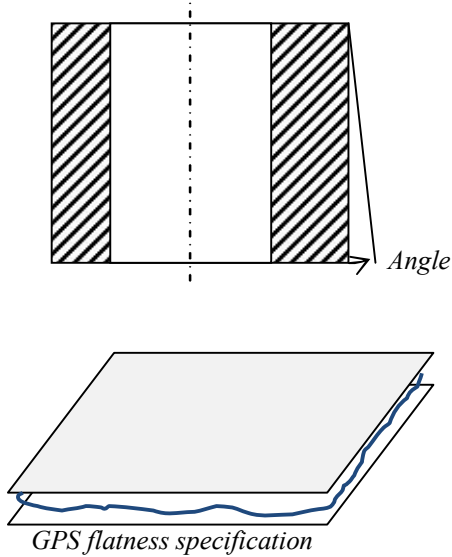

**Figure 4 :** *Parameters of the study: flatness and perpendicularity (angle)*

*3.1.2 Taguchi table*

The taguchi table we use (table 2), gives the order of the experiments and the levels we have to use for each one.

|   | A | B | C |
|---|---|---|---|
| 1 | 0,03 mm | 0,03 mm | 0,1 ° |
| 2 | 0,03 mm | 0,006 mm | 0,02 ° |
| 3 | 0,03 mm | 0,0015 mm | 1 ° |
| 4 | 0,006 mm | 0,03 mm | 0,02 ° |
| 5 | 0,006 mm | 0,006 mm | 1 ° |
| 6 | 0,006 mm | 0,0015 mm | 0,1 ° |
| 7 | 0,0015 mm | 0,03 mm | 1 ° |
| 8 | 0,0015 mm | 0,006 mm | 0,1 ° |
| 9 | 0,0015 mm | 0,0015 mm | 0,02 ° |

| Experiments for L9 | Number of parameters |   |   |   |
|---|---|---|---|---|
|   | 1 | 2 | 3 | 4 |
| 1 | 1 | 1 | 1 | 1 |
| 2 | 1 | 2 | 2 | 2 |
| 3 | 1 | 3 | 3 | 3 |
| 4 | 2 | 1 | 2 | 3 |
| 5 | 2 | 2 | 3 | 1 |
| 6 | 2 | 3 | 1 | 2 |
| 7 | 3 | 1 | 3 | 2 |
| 8 | 3 | 2 | 1 | 3 |
| 9 | 3 | 3 | 2 | 1 |

**Table 2:** *L9 design of experiment parameters*

According to the values of the Taguchi table, we have to construct 9 virtual parts.

**3.2 The building of a virtual part**

The part we use needs to define 6 faces and a cylinder. To build a model for a virtual part [15], we use a CMM to measure a real surface. We obtain the coordinates of 25 points for the plane. We also measured a real cylinder to have a set of points for calculation.

The second step of the preparation of the virtual part is to use an Excel sheet to generate a theoretical set of points. We have done 3 quality for the planes, so we have 3 sets of 25 points. We do not use directly the XYZ measured coordinates. We have to apply the least square method [13] to obtain an optimized plane with a theoretical normal direction (0,0,1) with a Z=0 position.

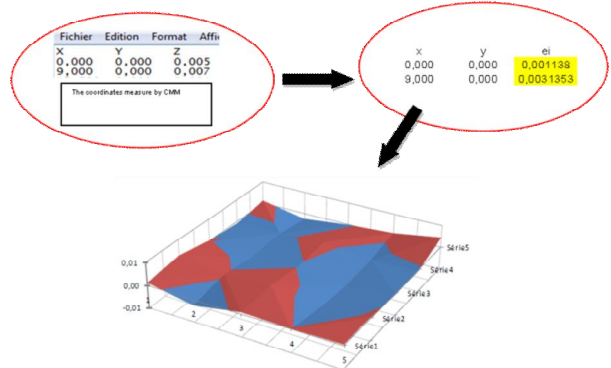

$$e_i = z_i - \alpha . y_i + \beta . x_i - \omega$$

**Figure 5:** *Least square method application*

Before to import the set of points (coordinates of the model) in CATIA, we have to save files in text format, with an adequate format. For each quality for the plane surface,

11

we obtain a set of 25 points: (Xi, Yi, ei) (as shown on fig. 5).

In the reality, the two planes (top and side) are perpendicular to each other with an "error" due to manufacturing. Because it is difficult to machine an exact default, we use virtual part. To obtain a perpendicularity default between top and side plane, we use CATIA to generate a set of point with the right position. CATIA software helps us do this easily by a simple rotation of the plane and an additional translation if needed [4]. Within the limits of this paper, we created three virtual perpendicularity defaults with details of plane angles: 90.002, 90.1, and 91 degree. We have 0.02, 0.1, and 1° for the angle parameter. The process is shown in fig 6.

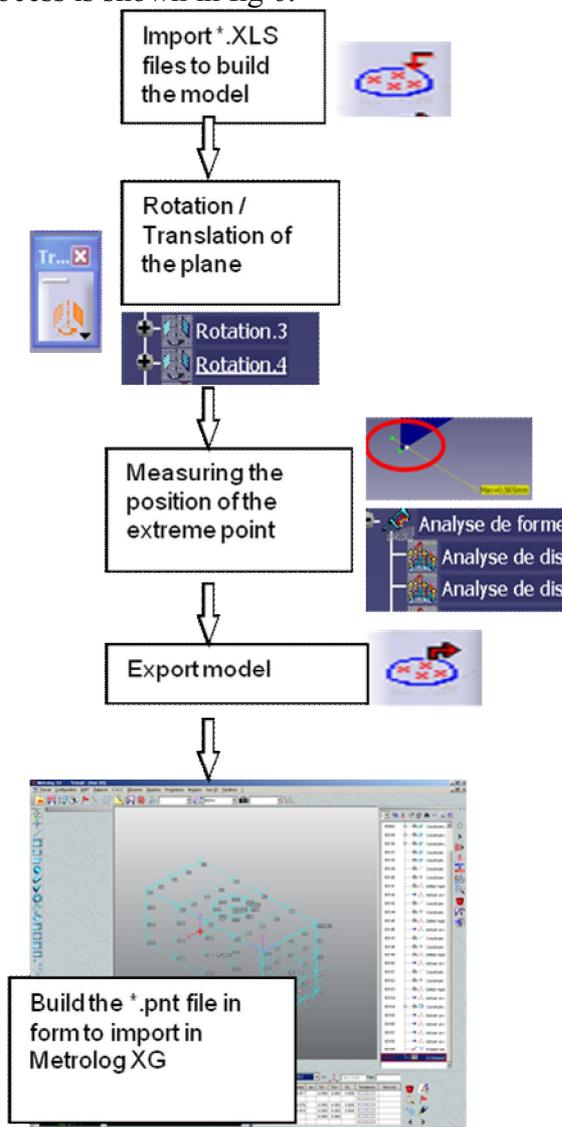

**Figure 6:** *Import/export from CATIA to METROLOG*

At the end of the virtual parts building process, we have 9 sets of points. Each set is in conformity with an experiment of the Taguchi table. For this first paper on the subject, the parameters are limited to 3 levels. Later, we can imagine increasing the number of experiments.

### 3.3 The calculation method

The part we use for the experiment permit to define a specification of position of a hole. It is a question of determining which limits of the defects of form involve a variation on the measure of location. The main feature of the specification of position is to imply the construction of a reference coordinate system. Various datum systems can exist: 3 planes and plane-line-point are the most used. In our case, we use a Plane/Line/Point system of coordinate.

The options included in the Metrolog software allow defining four variations for the coordinate system. We can mix the options to obtain different systems: tangent association between skin model and theoretical plane or least square method; simple intersection or plane with perpendicularity constraint.

The results of the four possibilities are shown on the figure 7 above.

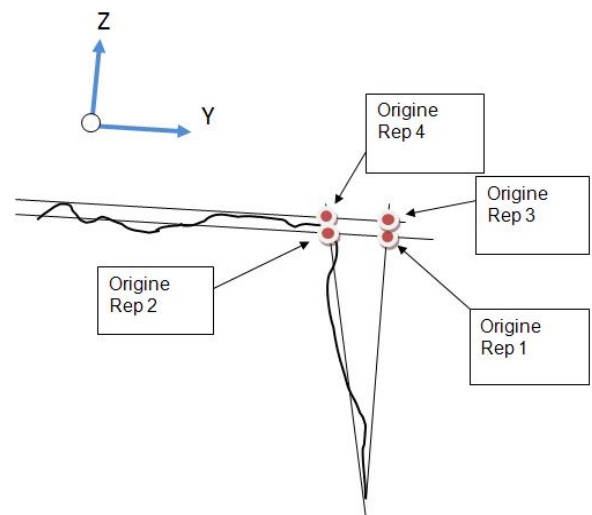

|  |  | Rep 1 | Rep 2 | Rep 3 | Rep 4 |
|---|---|---|---|---|---|
| **Top plane** | Least square | x | x | | |
|  | Tangent | | | x | x |
| **Side Plane** | Least square | | x | x | x |
|  | Tangent | x | | | |
| **Intersection** | | | | x | x |
| **Constraint of perpendicularity** | | | x | x | |

**Figure 7:** *Different models for the association*

We decide to limit the calculation to only one direction: the Y axis. The Rep 2 is the more simple to obtain. It is defined by two least square associations and an intersection to calculate the origin of the datum system. This is the "without-option" construction, and often, a non specialist user will take this one. We choose to take this datum system as the zero point for each part of the experiment. The value we can study is so the distance between Rep2 – Rep i with i in (1, 3 and 4).

It is important to notice that the best construction to be conforming to the norms is the Rep 3, with two tangent planes and a constraint of perpendicularity between these two planes. In the GPS (Geometrical Product Specification) norms, the theoretical planes associated to the surfaces A and B need to be tangent to the surfaces. In add, for the position specification ⌖ |A|B|, the norm define the construction of the datum coordinate system with a perpendicularity constraint between B and A (planes are defined on fig. 1).

### 4 Results and analysis

The dispersion on the distance between the datum system Rep2 and each Rep (i) possible to construct, gives some results.

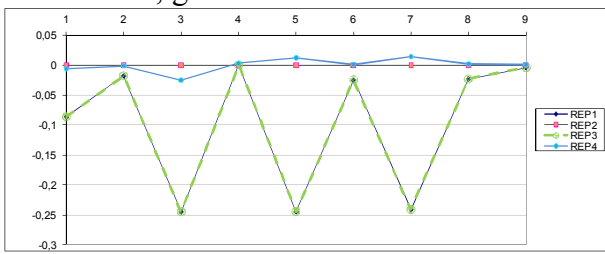

**Figure 8:** *The 9 experiments results*

The figure 8 shows that, if we used REP2 or REP4, the results in the 9 locations of the cylinder during experiments we perform involves quite the same results (small variations). This means that REP4 and REP2 are very similar but not totally identical.

The Rep3 and the Rep1 gives exactly the same dispersion. Because Rep3 is more conform to the definition of the norm, we will forgot Rep1 and only study the Rep3 and the Rep4.

For the Rep3, we can observe (see fig. 8) that 3 values have very important level: experiments 3, 5 and 7. These are the 3 experiments with the maximum value for the perpendicularity default: 1°. The 0,25 mm difference between Rep2 (simplest) and Rep3 (norm) is almost due to the angle. The length of the part with an angle of 1° gives 0,5 mm error even with perfects planes (theoretical planes). It is obtained by the calculation: lengh x tan(angle) = 30 x tan(1°) ~ 0,5 mm.

The figure 9 indicates that the angle is the main parameter to have an effect on the result of the position of the datum system. The figure 10 indicates that there is no influent parameter when we use the Rep4.

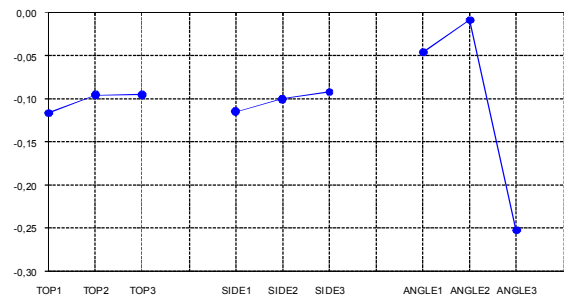

**Figure 9:** *Effects of the parameters on the Rep3*

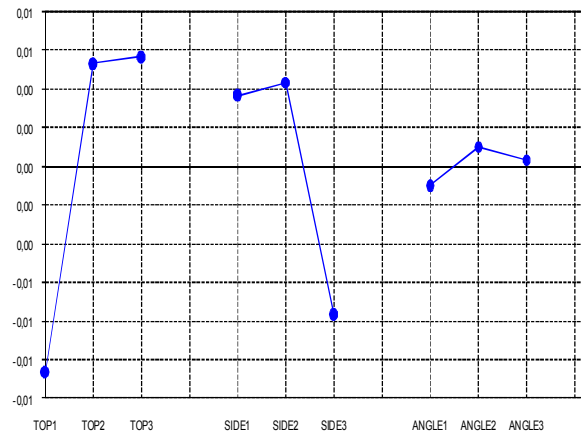

**Figure 10:** *Effects of the parameters on the Rep4*

In order to refine the result, we also check the dispersions with a view angle/angle. The table 3 gives some information about the form parameters without taking into account the angle.

| 0,02 ° angle | | | 0,1 ° angle | | | 1 ° angle | | |
|---|---|---|---|---|---|---|---|---|
| EXP2 | EXP4 | EXP9 | EXP1 | EXP6 | EXP8 | EXP3 | EXP5 | EXP7 |
| 0,02 | 0 | 0,01 | 0,09 | 0,03 | 0,03 | 0,25 | 0,26 | 0,26 |
| variation of position | | | variation of position | | | variation of position | | |
| 0,015 | | | 0,061 | | | 0,012 | | |

**Table 3:** *L9 design of experiment parameters for Rep3*

The values with the maximum angle are higher than with the other angles, but if we look only the 3 experiments with the maximum angle, the effect of the flatness of top and side planes is very low.

In the case of the medium value of the angle, the effect of the angle is almost 0,05 mm, and the variation of the result is more than this value. The flatness of the planes has an effect on the result of the position of a cylinder. We can do the same observation for the minimum angle.

When the perpendicularity specification is given between the planes A and B on a draft, the operator is able to be careful with this problem. But even if the effect is small, it must be taking into account by any operator who is making a measurement operation. The perpendicularity default always has an effect on the result of a position specification.

## 5 Conclusion

This article is a first step for a more global study. It turns out that the local coordinates of the parts when we build in the process of measurement is affected by the perpendicularity of the plane and when the angle is low, also affected by the flatness of the planes. These defaults in the construction method of the coordinate datum system are always present and any operator who is making the measurement of the position of a hole with a CMM must be aware of that.

We have used CATIA to construct the virtual parts for the experiment. Because it was a first study, we have limited the number of values for each parameter. With the first results, we can now define another study with more values. We hope that another design of experiment will confirm what we have find: the position of a cylinder in a 3D measurement is affected by the perpendicularity when this is the main default; it is affected also by the quality of the part surfaces (flatness of the top and side plane) when the angle is lower.